\documentclass{article}

\usepackage{arxiv}

\usepackage[utf8]{inputenc} 
\usepackage[T1]{fontenc}    
\usepackage{hyperref}       
\usepackage{url}            
\usepackage{booktabs}       
\usepackage{amsfonts}       
\usepackage{nicefrac}       
\usepackage{microtype}      
\usepackage{lipsum}
\usepackage{graphicx}
\graphicspath{ {./images/} }

\usepackage{tabulary}
\usepackage{tabularray}
\usepackage[normalem]{ulem}
\usepackage{multirow}
\usepackage{tablefootnote}
\usepackage{MnSymbol}
\usepackage{authblk} 

\title{State of the Art on Artificial Intelligence Resources for Interaction Media Design in Digital Cultural Heritage}

\author{
    {\bfseries Manuele Veggi} \\
    \begin{minipage}{0.45\textwidth}
        \centering
        Sapienza University of Rome\\
        Department of Science of Antiquities\\
        Rome, Italy 00185 \\
        \texttt{manuele.veggi@uniroma1.it}
    \end{minipage}
    \hfill
    \begin{minipage}{0.45\textwidth}
        \centering
        National Research Council\\
        Institute of Heritage Science, DHILab\\
        Florence, Italy 50144 \\
        \texttt{manueleveggi@cnr.it}
    \end{minipage}
}
\begin{document}
\maketitle
\begin{abstract}
This paper explores the integration of Artificial Intelligence (AI) in the design of interactive experiences for Cultural Heritage (CH). Previous studies indeed either miss to represent the specificity of the CH or mention possible tools without making a clear reference to a structured Interaction Design (IxD) workflow. The study also attempts to overcome one of the major limitations of traditional literature review, which may fail to capture proprietary tools whose release is rarely accompanied by academic publications. Besides the analysis of previous research, the study proposes a possible workflow for IxD in CH, subdivided into phases and tasks: for each of them, this paper proposes possible AI-based  tools that can support the activity of designers, curators, and CH professionals. The review concludes with a final section outlining future paths for research and development in this domain. 
\end{abstract}


\section{Introduction}
\label{a:introduction}

In recent years, the outstanding performances of Artificial Intelligence (AI) have had a considerable impact in multiple and heterogeneous fields of knowledge. In particular, a new impulse has been given to reshape research and dissemination in the domain of Cultural Heritage (CH). A number of studies, white papers, and briefs have already highlighted how AI can enhance the understanding, management and communication of CH. Izsak and colleagues mention as major areas of implementation archival, cataloguing, information management; visitor experience management; audience engagement activities \cite{izsakoppai}. Similarly, the European Parliament itself has highlighted how these classes of algorithms have already benefited scholars and cultural heritage professionals in (meta)data management, restoration, authorship attribution, and artworks detection and localisation \cite{magdalena2023artificial}. In addition, several experiments have demonstrated how AI offers unprecedented solutions in museums, with respect to curation, audience engagement, and content creation \cite{book}.

This continuously evolving landscape also provides a unique stimulus for User Experience (UX), interaction designers and curators in the CH domain: the implementation of AI techniques may considerably streamline the creative and design process and suggest novel interaction and engagement strategies in the final visitor experience. This paper is therefore based on the awareness of these ongoing challenges, and aims to provide a brief overview of some possible AI-based solutions for designers of interactive experiences to make a more effective use of CH. \footnote{The data collection stage of this research was concluded in February 2024. Resources and services released after this date are therefore not included in this study.}

Upon initial examination, it becomes evident that a number of significant difficulties exist that hinder the achievement of this objective. Firstly, the AI field is in a constant state of evolution, with new and more sophisticated classes of algorithms and engineering solutions being constantly developed. Furthermore, this ongoing research and development effort results in a vast array of resources becoming available. In order to provide a comprehensive overview of the current state of the art (SOTA), it is necessary to organise the available material in a systematic manner. To this end, a possible solution may be retrieved when considering Tom Mitchell’s definition of machine learning (ML), on which different classes of AI services are based: “a computer program is said to learn from experience (E) with respect to some class of tasks (T) and performance measure (P), if its performance at tasks in T, as measured by P, improves with experience E” \cite{awadMachineLearning2015}.

This seminal definition is insightful for this study, as it describes three different perspectives from which AI applications in UX Research and Design may be observed:

\begin{itemize}
    \item the \textit{experience}, i.e. the processes and training epochs which allows AI systems to effectively learn and improve their performance;
    \item the \textit{performance measure}, i.e. framework and evaluation tools to assess the performance of an AI algorithm and eventually identify gold standards technologies;
    \item the \textit{class of tasks}, i.e. the procedures which may be delegated - partially or totally - to an AI system.
\end{itemize}

The present study attempts to address this research problem embracing this third perspective. It is noteworthy that the majority of the presented solutions are proprietary, whereas open-source services are scarce. Consequently, it is challenging to have free access to both the model underlying the front-end AI-powered service and to the full set of functionalities of an application. Furthermore, as this study is intended for both academic and non-academic audiences, including designers, curators, and CH professionals who may not have specific AI expertise, providing an overview of the SOTA in terms of automating specific tasks that align with their daily work is more relevant with the requests and needs of this target audience.

Consequently, the research question this paper seeks to address is to determine which tasks might benefit from the implementation of AI copilot in the design process and to identify potential examples. The reuse of the “copilot" metaphor (taken from the GitHub Copilot service, see \textit{infra}) serves to highlight an essential methodological caveat: neither the performance quality in some task classes nor, above all, the complexity of Cultural Heritage justifies the delegation of the entire design process to non-human actors. It is evident that the pivotal role of human resources in the governance of AI systems applied to CH is irrefutable. This has been recently emphasised by the European Parliament \cite{magdalena2023artificial}; furthermore, the undeniable importance of human oversight of AI applications in this domain is also reflected in fundamental international charters on the integration of ICT for the exploitation and visualisation of CH, such as the London Charter \cite{denardLondonCharterComputerBased2009} and the Seville Principles \cite{icomosPrinciplesSevilleInternational2017}.

The SOTA overview will be structured as follows: after an initial survey of previous studies on the topic, the adopted methodology will be discussed in the third section of this paper. This paragraph will also organise the “class of tasks” in which AI may be applied in a structured workflow for IxD in the CH domain. This lays the basis of the central section of this work, where, for each task, possible existing services are mentioned. The results of this overview will later be discussed, and future lines of research will be proposed in the conclusion.

\section{Previous Studies and Existing Literature}
\label{a:sota}

A rapid overview of online publication databases immediately highlights two major problems in the existing literature about this research field. The first one pertains to its content and is mainly due to the fact that the current research question addresses a deeply interdisciplinary field, which may be understood as an intersection of AI applications in both CH (Domain 1) and UX / IxD (Domain 2). As already seen in the introduction, scholars and public bodies are today well aware of the opportunities and risks of the former domain and a systematic overview of this topic would exceed the scope of the current research. On the other hand, a preliminary discrimination is necessary when analysing publications covering the relationship between AI and UX Design: indeed, part of them focuses on how UX may benefit and enhance human access to AI (\textit{UX for AI}) (as an example, see \cite{lewAIUXWhy2020}).

\begin{figure}[htb]
  \centering
  \includegraphics[width=.8\linewidth]{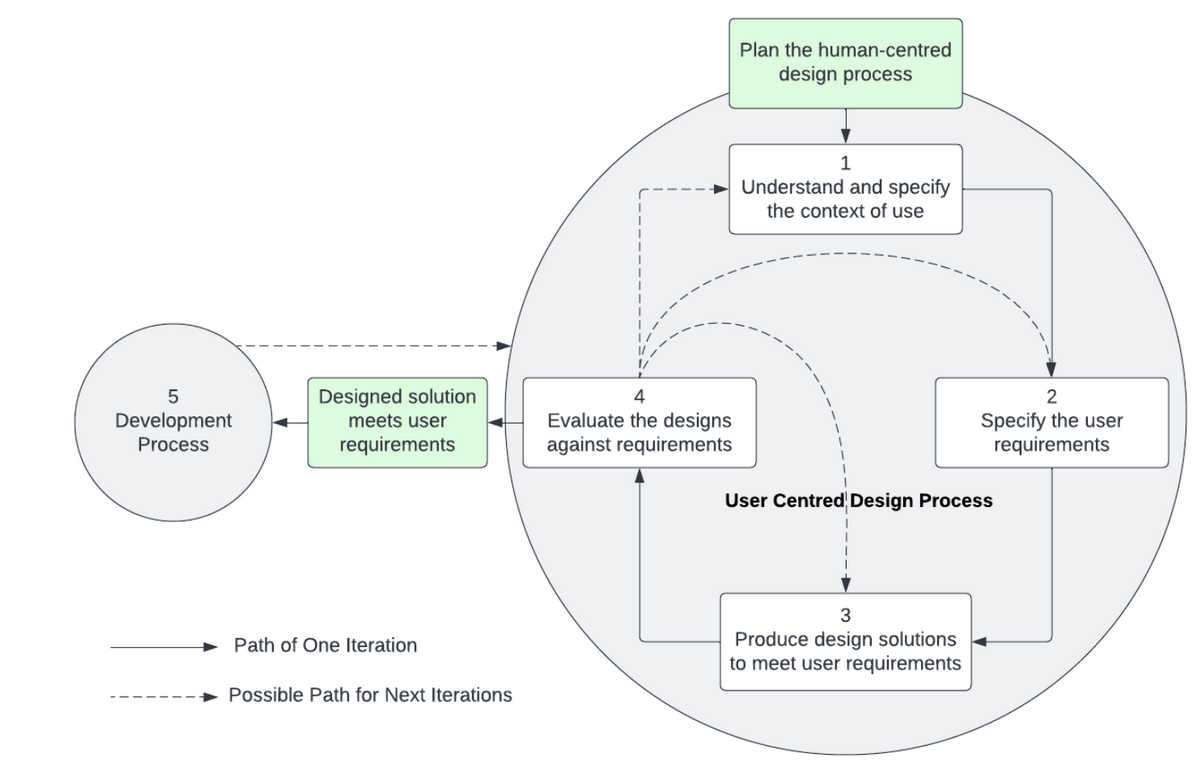}
  \caption{\label{fig:img1}Workflow for Human-Centred Design \cite{stigeArtificialIntelligenceAI2023}}
\end{figure}

As a result, scholarly contributions in Domain 2 on \textit{AI for UX} are limited, yet some relevant studies may be found. To the best of my knowledge, one of the most complete and recent overviews has been provided by Stige and colleagues \cite{stigeArtificialIntelligenceAI2023}. Here, the authors performed a robust and systematic literature review, based on 46 items (journal papers, conference proceedings, or book chapters) published after 2015. Their research focuses on a human-centred design workflow, based on ISO standard 9241-210 (see Figure \ref{fig:img1}): this approach aims to “make [interactive] systems usable and useful by focusing on the users, their needs and requirements, and by applying human factors/ergonomics, and usability knowledge and techniques" \cite{iso}. After the presentation of this design methodology, the study identifies possible areas of implementation of AI techniques through a qualitative analysis of the publications in the retrieved corpus: digital design process, understanding of the context of use, user requirements specification, solution design, and design evaluation.

A comparable outcome has been achieved by Virvou \cite{virvouArtificialIntelligenceUser2023}. The author identifies a number of different classes of UX which may benefit from the implementation of AI techniques. These include web search, recommender systems, intelligent help systems and virtual assistants, and intelligent tutoring systems. For each of them, the author also proposes relevant classes of technology as effective examples: they range from ML to knowledge graphs, including also Natural Language Processing (NLP) and Generative AI (GenAI). The author posits that these tools facilitate the collection of more precise information about the target audience, thereby enabling the design of more tailored and personalised UX and UI solutions.

In this context, particular attention is directed towards the role of GenAI in the field of UX and UI design. For example, an experimental study by IBM demonstrated how the redesign of interactive applications and the UX modernisation processes are enhanced and streamlined by these technologies \cite{Houde2022OpportunitiesFG}. Specifically, the author “identified a number of pain points within the UX modernisation process [...] that could potentially be addressed by new generative AI technology in the form of a set of generative AI models” \cite{Houde2022OpportunitiesFG}. These narrow AI systems (i.e. AI for dedicated purposes) would assist designers  with multiple tasks, such as modernisation planning, design specification and implementation. Finally, other studies have been conducted on the potential integration of AI methods in the design and development of video games \cite{westeraArtificialIntelligenceMoving2020}.

\begin{figure}[htb]
  \centering
  \includegraphics[width=.8\linewidth]{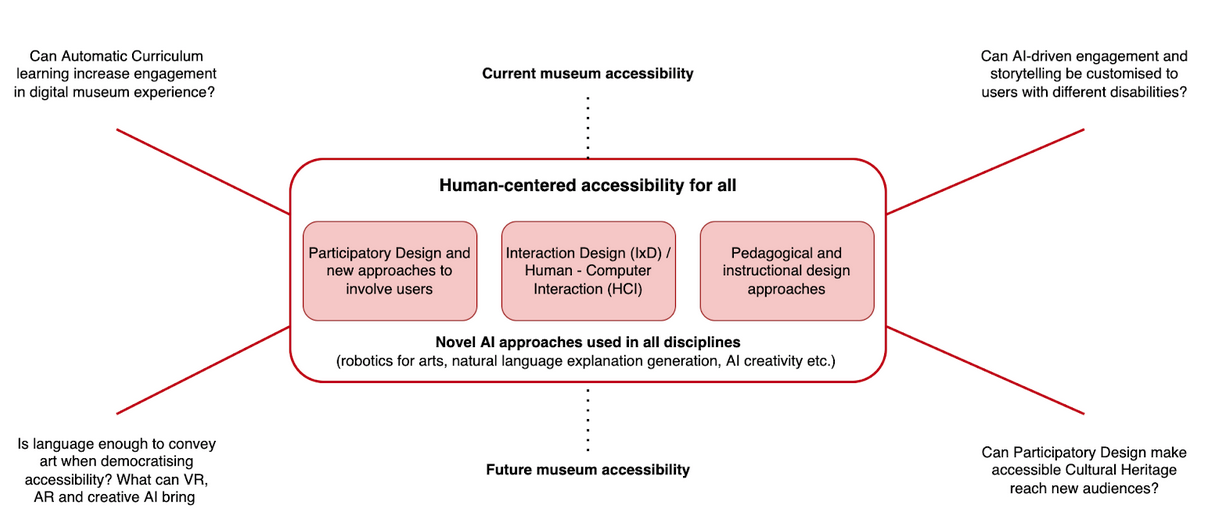}
  \caption{\label{fig:img2}Conceptual framework for AI-enabled accessibility of museums and cultural heritage sites \cite{pisoniHumanCenteredArtificialIntelligence2021}}
\end{figure}

A search of bibliographic resources yielded only one SOTA review on AI for UX, with a core focus on CH. In 2021, Pisoni and colleagues published an extensive overview of the topic, differentiating among different areas of expertise (interaction design, pedagogical and participatory design) and demonstrating how AI offers unprecedented solutions for the accessibility of CH collections \cite{pisoniHumanCenteredArtificialIntelligence2021}. An essential step in their research is the proposal of the following conceptual framework, in which classes of technologies are associated with their purpose. This model identifies three main use cases of AI: “AI, Creativity, Storytelling and Audience Engagement”, “Explaining Art through Language”, and “Art and Robotics”. The literature review showed the predominance of Mobile Remote Presence (MRP) systems and, again, NLP and GenAI (see Figure \ref{fig:img2}).

Nevertheless, the promising results of this analysis do not encompass the entirety of the design workflow. Instead, the authors describe the final products and potential use cases. Similarly, the SOTA reviews of Domain 2 fail to capture the specificity of digital cultural heritage. As an example, the proposal by Stige et al. in the phase of solutions design primarily concentrate on UI. However, there appears to be a lack of consideration given to the specific types of assets that may be encountered in an interactive media application for CH, such as 3D models, narrative frameworks (storytelling), music, and so forth.

Moreover, an additional problem with the existing literature is related to the methodology used. Indeed, these state-of-the-art overviews are based on corpora of academic publications, as exemplified by the \textit{Guidance on the Conduct of Narrative Synthesis in Systematic Review }\cite{popay2006guidance}, which serves as the foundation for Stige and colleagues’ overview. This rigorous and promising approach nevertheless does not cover several contributions to Domain 2, which are developed by private software companies and whose release may not be accompanied by scientific publications. This peculiarity of the research domains also raises significant methodological questions, which will be addressed in the following section.

\section{Methodology}
\label{a:methodology}

The studies referenced in Section \ref{a:sota} employ an inductive reasoning methodology. This entails querying a general publication database and, through the definition of clear inclusion criteria, deriving a reference corpus. The analysis of the individual resources then provides a general overview. This approach is rigorous, yet it excludes a priori services and resources that are not described by scholarly publications, such as the products of private software companies. Nevertheless, their analysis is indispensable for this research, which targets a diverse audience of academic researchers, CH professionals in public institutions, and corporate employees.

The absence of a comprehensive catalogue of AI services in UX for CH precludes the possibility of replicating a similar inductive approach. Conversely, a deductive reasoning approach may be a valid methodological framework for the purpose of this analysis. Consequently, this research will commence with the definition of a general workflow for IxD in the CH, which will be organised into core tasks and possible subtasks (see Section \ref{a:workflow}). Each of these processes will be then considered, and the potential applications of AI algorithms or resources will be described (see Section \ref{a:list-resources}). In the absence of a reference catalogue, these digital products will be retrieved through web search.

Furthermore, the difficulty of carrying out a comprehensive recognition and a systematic assessment of the available AI technologies in this field (see \textit{supra}) complicates the definition of inclusion criteria for this analysis. Consequently, the objective of this work is not the creation of an extensive catalogue of available resources. Nor the inclusion or exclusion of a specific product should  be interpreted as an implicit judgement of its performance. Rather, it should be considered as a mere exemplification of a possible use case of an AI technique for a specific task in the IxD workflow. Section \ref{a:ai-survey} provides a preliminary consideration of how AI tools are effectively used in everyday working practice by designers, curators and CH professionals and researchers.

\subsection{A Possible Workflow for Interaction Design in Digital Cultural Heritage}
\label{a:workflow}

The objective of this section is to define an IxD workflow for CH. In the aforementioned deductive methodological perspective, this model is essential as it summarises all the different tasks and subtasks which are involved in the life cycle of an interactive application in this domain. The field of UX and IxD has already provided a solid foundation for the development of this model. As illustrated in Figure 1, Stiege and colleagues proposed a potential workflow, yet the individual steps often fail to fully capture the distinctive characteristics of those design practices that are particularly relevant to the heritage domain. Consequently, the model presented in this paper draws upon the most effective practices within this sector, reconstructed using some of the most authoritative handbooks, such as \textit{About Face} by Alan Cooper \cite{cooperFaceEssentialsInteraction2007} and David Benyon’s \textit{Designing Interactive System} \cite{benyonDesigningInteractiveSystems2013}. 

At this juncture of the discussion, it becomes crucial to acknowledge the nuanced nature of the Cultural Heritage domain, and the limited number of resources available in this particular subfield. Upon closer examination, it becomes evident that the majority of published articles on the topic either focus on a potential workflow for visitor experiences that rely on a specific class of technologies, as exemplified by the methodology proposed by Cipriani and colleagues on reality-based 3D models \cite{isprs-archives-XLII-2-W11-427-2019}, or describe the process for developing a single visitor experience, without providing a general abstract workflow for other case studies. A potential exception to this dichotomy can be observed in a co-design tool, \textit{VisitorBox} (\url{https://visitorbox.wp.horizon.ac.uk/}), created by a team from the University of Nottingham \cite{darzentasDatainspiredCodesignMuseum2022} within the context of the Project GIFT (\url{https://gifting.digital/}). The toolbox comprises a printable set of cards that assists visitor experience designers in defining the requirements of the star assets, the target audiences, and the possible interaction phases of a digital experience for cultural institutions. However, it does not provide methodological guidelines for developing, testing, and redesigning the final product.

\begin{figure}[htb]
  \centering
  \includegraphics[width=.8\linewidth]{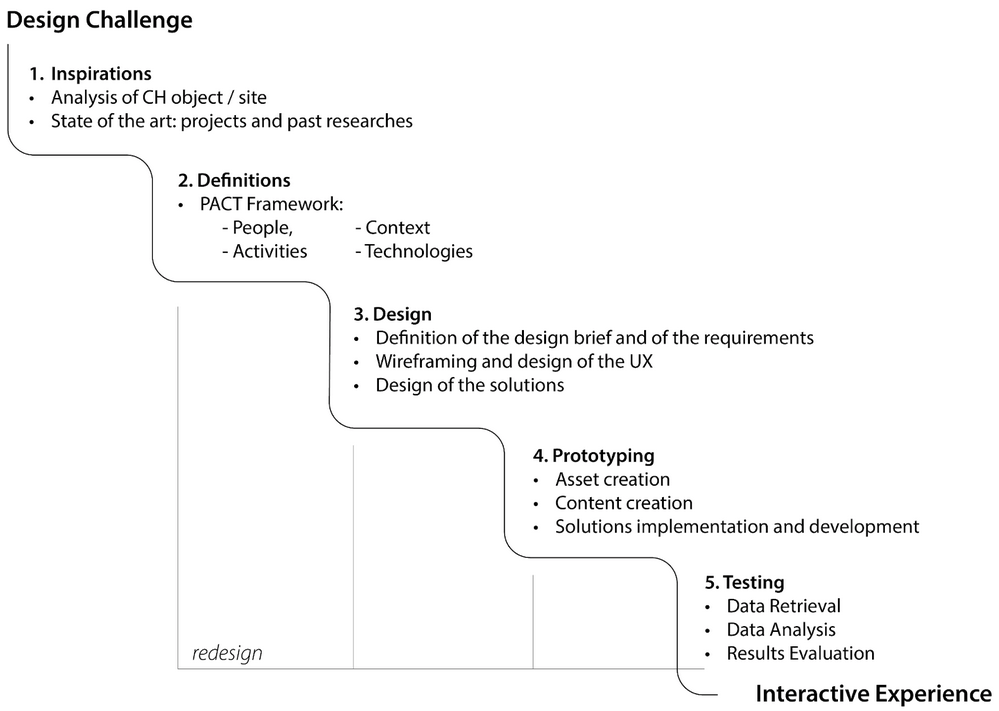}
  \caption{\label{fig:img3}Interaction Design Workflow for the development of cultural heritage experiences}
\end{figure}

This paper therefore proposes a possible pipeline for the development of CH visitor applications, relying on the best practices and guidelines described above. It also draws upon the experience of the Italian National Research Council - Institute of Heritage Science (CNR - ISPC), which is the principal stakeholder of this study and which has been developing interactive digital applications for CH for several years, also in the context of national and European projects. A preliminary overview of this workflow is presented in Figure \ref{fig:img3}. The subsequent description of the various stages of this pipeline outlines a general set of tasks that may be necessary for the development of different applications. Nevertheless, it is possible to identify further potential sub-steps that may vary considerably from the typology, features and purposes of the final application. These are thus excluded from the current analysis.

This workflow can be readily integrated with the more renowned Double Diamond model \cite{designcouncilDoubleDiamondUniversallyn.d.}, which suggests that designers should initially \textit{discover} the domain in which they are working. In the CH context, this initially entails a comprehensive examination of the distinctive characteristics of the selected case study, drawing upon a diverse array of insights from various disciplines, including art history and conservation science. This step should be conducted in the broader framework of a state-of-the-art review of contributions which will support designers and curators in a better comprehension of several features of the final products. These may include existing literature on the target audience and on the cognitive goal. Furthermore, it is advisable to consider past projects and applications in order to gain insights that can inform the entire design process. Given the preliminary nature of this multiple research, this stage has been designated as “inspirations”, as it provides the necessary suggestions to initiate the creation of the final visitor experience.

The second step of the Double Diamond, \textit{define}, encompasses the activities mentioned in the second and the third phases of the proposed workflow. The first foresees the definition of the main aspect of the application after the PACT Framework proposed by Benyon \cite{benyonDesigningInteractiveSystems2013}. This model demonstrates how the design of an application should be informed by an in-depth analysis of four core aspects involved in the UX: the people, considering their physical, psychological, social differences and their mental model; the actions, with a particular focus on the temporal aspects. Additionally, the possibility of collaboration, the level of complexity and the safety criticality, as well as the nature of the content; the physical environment and the social and organisational context; and the technologies involved, were analysed in terms of inputs, outputs, communication processes and provided content. A scrupulous application of this framework will lead to “the best combination of the PACT elements with respect to a particular domain, [getting] the right mix of technologies to support the activities being undertaken by people in different contexts” \cite{benyonDesigningInteractiveSystems2013}. 

The subsequent stage of the workflow, design, deals with the formulation of the design for the final application. This phase, which can be carried out with the assistance of (co-)design tools such as the aforementioned \textit{VisitorBox}, primarily coincides with the preparation of the design brief, defined as “a thorough presentation of the problem, together with the expected outcomes of design”. This technical document should address fundamental questions pertaining to the final application, including the rationale behind the project, the expected outcomes, the target audience and key stakeholders, the available time and economic resources, and the assessment of the results obtained \cite{sasEnhancingCreativityInteraction2009}. The brief should in addition include a definition of the design solutions and their final development, including a generic description of the main metaphor of the experience and of the storytelling, together with the first wireframes of the UI of the final applications. This stage is of critical importance, as it paves the way for all the necessary actions to be carried out in the fourth phase of the workflow (“prototyping”).

In the Double Diamond, the prototyping phase is largely concurrent with the third section, i.e. \textit{develop}. This deals with the generation of the various assets comprising the application and the requisite content to be displayed in the final virtual experience. More specifically, this encompasses the preparation of audio, video, image, 3D object and environment, and text assets. This phase also involves the development of the entire software architecture, including animations and the designed interactions. Finally, the fifth step pertains to the testing of the final visitor experience. In this phase, designers should prepare an evaluation protocol to collect data about users’ reactions to the current version of the interactive application. The collected data set should then be analysed in order to inform the subsequent redesign of the pain point of the visit experience. The implementation of the correction and the creation of the initial version of the User Experience (UX) coincides with the final step of the Double Diamond model, labelled as \textit{deliver}.

\subsection{AI in Interaction Designers’ Everyday Practice}
\label{a:ai-survey}

Scholars have only rarely focused on surveying the current state of implementation of AI tools in UI and UX designers’ working practices. Furthermore, the bibliographic research returned no contribution specifically focused on the CH. Indeed, similar overviews can commonly be found on design blogs and communities, often lacking a clear methodological statement on how data is acquired, nor providing information on the population involved in the data collection process. As an illustration, a Hubspot Blogs Research study conducted an interview in the web design domain and discovered that more than nine out of ten designers are utilising AI tools in their design practice, with over half of the participants indicating that their company encourages the use of similar technologies \cite{vettorinoHowWebDesigners2023}. The analysis revealed that the most common areas of implementation are image and media asset generation (58\%), complete web page design creation (50\%), testing of new design strategies (49\%), suggestion of possible improvements in the design (43\%), tracking of a specific design performance or quality (40\%), and auditing of user experience (20\%).

These results appear to be at odds with the findings of a more comprehensive analysis conducted in 2021 on over a hundred UX/UI practitioners in Brazil \cite{bertaoArtificialIntelligenceUX2021}. In this study, only a minority of designers reported integrating AI into their daily work. It is important to note that the survey was conducted two years before the release of ChatGPT 3 and the subsequent “hype" surrounding this class of GenAI (for further insight, see \cite{heppChatGPTLaMDAHype2023} and \cite{rooseHowChatGPTKicked2023}). This may be a contributing factor to the results of this research. In fact, the analysis revealed a markedly different scenario than what is currently observed in designers' practices. Only 21\% of respondents indicated that they had developed projects supported by AI, and less than one participant out of five claimed to have used AI-powered design, such as image background removal. Only an extremely scarce minority of the population (5\%) used AI-integrated design systems, such as Adobe Sensei.

\begin{figure}[htb]
  \centering
  \includegraphics[width=.8\linewidth]{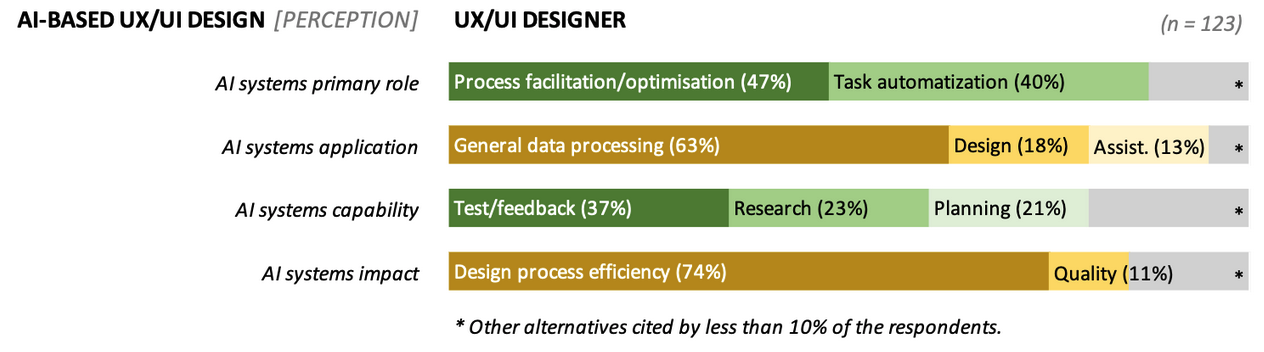}
  \caption{\label{fig:img4}Perspectives on AI-based UX/UI design \cite{bertaoArtificialIntelligenceUX2021}}
\end{figure}

In contrast, the \textit{perception} of the possible fields of implementation of this class of technology was more optimistic. Figure \ref{fig:img4} indicates that respondents to the survey perceived AI as a potential tool to support design practices. In another question, the results demonstrate that the population viewed these technologies as either a “co-creator" (41\%) or a “collaborator" (26\%) in the future. The anticipated use case primarily involves the automation of tasks, process streamlining, data processing, and efficiency enhancement.

These studies describe a scenario that is undergoing rapid change, with consistent variations across countries. The latter study does, in fact, observe significant differences with the results of other studies in other geopolitical contexts. The findings of Bertão and Joo diverge from those of other research “in the US, UK, and Scandinavia, where 63\% of the respondents claim to have worked with AI” \cite{bertaoArtificialIntelligenceUX2021}. The authors cite a previous study conducted in 2017 by Dove and colleagues, which also found that AI tools were an effective solution for collaboration in the design process. In particular, the study highlighted the potential of AI tools for “making sense of user actions, [...] [in]  the form of recommendation engines, and fraud detection system[s]” \cite{doveUXDesignInnovation2017}.

In conclusion, these results demonstrated a significant shift in design practices and current AI tool implementation, while also indicating a shared perception of their potential to streamline, optimise, and enhance the design process. This consensus therefore provides the impetus for the current study, which is aimed at delineating the current state of the art on the potentialities of AI as a collaborative tool for IxD and UX design, with a particular focus on CH.

\section{AI Techniques in Interaction Design for Cultural Heritage}
\label{a:list-resources}

\subsection{Phase 1: Inspirations}
\label{a:phase1}

As described in the methodology section, in the first phase of their workflow designers scrutinise the reference domain of the project, taking into account existing resources. Traditional approaches of market analysis are yet difficult to be applied to Digital Cultural Heritage (DCH): resources are not always well documented, and their sustainability is often not considered. Moreover, to better exploit their collections, public institutions collaborate with universities and research centres, describing these new visitor experiences also with scholarly publications. Compared to a mere web research, bibliographic research manages to capture in a more consistent manner the existing products and may provide useful insights to guide the design process (analysis of the artwork / CH site, literature describing the final goal of the application, etc.).

As the current study is not aimed exclusively at cultural heritage scholars, the “[incorporation of] AI systems in the understanding of tangible, documentary and intangible cultural heritage", which are welcomed in the UNESCO Recommendation on the Ethics of Artificial Intelligence  \cite{unescoRecommendationEthicsArtificial2022}, are not described in this section. A complete overview of these methods, which are traditionally used in digital art history and other subfields of the digital humanities (for a first introduction see \cite{brownRoutledgeCompanionDigital2020}), would go beyond the scope of the research question and the target audience of this study.

Nevertheless, these analyses are carried out by academics and scholars and their results are often disseminated through bibliographic products. As a result, secondary research on journal articles, books or proceedings allows for the collection of multiple, scientifically validated and therefore reliable data on the CH object or site chosen as the main case study for the final visitor experience. In addition, in the case of IxD applied to Digital Cultural Heritage, this research is not limited to the “understanding” of the work of art, but also aims to retrieve pieces of information about the goal of the final application, the design methodology, possible technologies to develop the UX, as well as potential tools and approaches to carry out the evaluation frameworks.

Therefore, this section focuses on “narrow AI" systems to identify the state of the art based on academic publications, which includes two main activities. On the one hand, a researcher needs to scrutinise all the different resources and try to gain a general overview of the chosen topic; on the other one, it is also necessary to analyse in detail the single bibliographic item. Possible AI tools to perform these tasks are described in the two following subsections.  

\subsubsection{Task 1.1: Identification of the Existing Resources}
\label{a:phase1-1}

In recent years, different tools have been released to streamline the process of researching contributions to draw the state of the art on a specific domain. One of the best known examples is the database of Google Scholar (\url{https://scholar.google.com/}), which allows users to query several publications through specific keywords and to activate filters to refine the output of the search. These functionalities have been enhanced by the Semantic Scholar service (\url{https://www.semanticscholar.org/}): here, the users’ research is improved by the implementation of AI tools to enhance their query , for instance by indicating most influential contributions, by offering users a system of paper recommendations, as well as the possibility to integrate a very short summary of a contributions to speed up the initial exploratory bibliographical search on a topic through the \textit{Too Long; Didn't Read }(TLDR) service.

This core set of functionalities is incorporated in a huge variety of services, often based on NLP techniques. Consensus (\url{https://consensus.app/}) is able to reply to user-generated questions, on the basis of existing research papers, providing a brief summary of the state of the art and listing the most important contributions, together with their metadata. This solution marks a significant improvement compared to ChatGPT, whose reliability about cited references has been often questioned \cite{frosoliniReferenceRoleChat2023}. Similarly, Scite (\url{https://scite.ai/}) is capable of answering a question asked by the user: the outcome of this interaction is a brief text in which the system summarises the findings of the most relevant publications and, for each contribution, the most relevant metadata are shown, as well as the possibility to export the citation into different format. In addition, the entirety of the existing contributions may be visualised as graphs, in order to visually infer which articles may be most influential. Together with Scite, also Elicit (\url{https://elicit.com/}) implements AI techniques to retrieve most relevant pieces of information from a paper and return to the user a synoptic overview of the state of the art of the domain. Users of the platform may perform a keyword search and, in response, Elicit provides a brief summary, mentioning the most relevant articles on the topic. In addition, it shows a “list of concepts” associated with the input keyword(s) and the corresponding publications where the topic may be better scrutinised. 

\subsubsection{Task 1.2: Single Source Analysis}
\label{a:phase1-2}

While the resources defined in Section \ref{a:phase1-1}. may support the reconstruction of the state of the art by comparing simultaneously different scholarly publications, this paragraph focuses on the opposite task for literature review, i.e. the in-depth analysis of the single bibliographic item. Semantic Scholars and other services already provide simple summaries for every article (see \textit{supra}), yet more specific AI-powered solutions can help in skimming, summarising and retrieving the most relevant information. For the first task, Semantic Scholar offers also a specific service: \textit{Skimming Assist} is an embedded functionality in the PDF viewer (\textit{Semantic Reader}) that highlights the most relevant aspects of the paper, adding notes in the margin. As a result, users can directly identify goals, methods and results described in the scholarly contributions. 
More versatile tools allow also to manipulate non-academic research, providing summaries and allowing readers to directly interrogate specific documents. In particular, ChatPDF (\url{https://www.chatpdf.com/}) well exemplifies this latter task: on this platform it is possible to directly upload a PDF document, which provides the knowledge base for a chatbot which can be directly interrogated by the user. Specific questions on the content of the document and the system will provide an answer, citing the pages from which this information was extracted. 

\subsection{Phase 2: PACT Framework}
\label{a:phase2}

As anticipated in Section \ref{a:workflow}, the definition of the requirements and of the main structure of the interaction can be supported by the implementation of Benyon’s PACT Framework \cite{benyonDesigningInteractiveSystems2013}. The majority of the AI technology in this sector yet seems to provide a more relevant contribution in the analysis, understanding and modelling only of the target audience. Multiple research identified how this task is crucial for any effective design process: in particular in the IxD domain, this step allows to design practitioners, through different methods, to “condense, organise, and clarify research data into a coherent vision of users and their goals; allow that vision to be communicated compellingly; to everyone with a stake in making the product a success objectify all key assumptions affecting design so they can be validated through research and discussion” \cite{daytonAudiencesInvolvedImagined2003}. The following section hence mentions possible solutions to automatise two of the most common tasks in UX Research: understanding and the modelling of the audience.

\subsubsection{Task 2.1: Understanding the User}
\label{a:phase2-1}

This task encompasses a series of activities and methodologies which characterise traditional UX Research: as underlined by Cooper \cite{cooperFaceEssentialsInteraction2007}, this procedure requires qualitative research techniques, which helps understanding behaviours and attitude of potential users, as well as the domain in which this product should be used. Some of the possible approaches traditionally adopted by UX Research are interviews with stakeholders, subject matter experts or potential users, user observations, and literature reviews \cite{cooperFaceEssentialsInteraction2007}.

Artificial Intelligence supports qualitative research mainly through natural language processing techniques. Different services such as Grain (\url{https://grain.com/}), Loopanel (\url{https://www.looppanel.com/}) and QoQo (\url{https://qoqo.ai/}) offer the possibility to automatically transcribe the interviews (speech recognition) into a searchable text, whereas Kraftful (\url{https://www.kraftful.com/}) is able to analyse user feedbacks from different sources (interview transcripts or comments on applications marketplace (AppStore or PlayStore), in order to identify the most critical pain point of the final product. 

Furthermore, these features are often framed in more articulated platforms, which are designed as co-working and project organisation tools and where the data retrieved during the qualitative research may be better organised. AI indeed provides an improved system for note taking during user observation: notes may be labelled and clusters created through similar notes. As an example, Notably.AI (\url{https://www.notably.ai/}) enhances this functionality by implementing a sentiment-based labelling of the notes or of the interview transcriptions. 

\subsubsection{Task 2.2: Modelling the User}
\label{a:phase2-2}

The majority of AI-powered tools support designers in users’ modelling by the definition of personas. This method is widely known in UX praxis and indicates “models of users based on observed behaviours and intuitive synthesis of the patterns in the data. Only after we formalise such patterns can we hope to systematically construct pat- terns of interaction that smoothly match the behaviour patterns, mental models, and goals of users. Personas provide this formalisation” \cite{cooperFaceEssentialsInteraction2007}.

Platforms such as UserDocAI (\url{https://userdoc.fyi/}) and the aforementioned QoQo, (based on OpenAI technologies and accessible as Figma plug-in) are able to create user personas starting from an essential description: the latter service, for instance, is able to define goals, needs, motivations, frustrations, tasks and opportunities of the potential users.

\begin{figure}[htb]
  \centering
  \includegraphics[width=.8\linewidth]{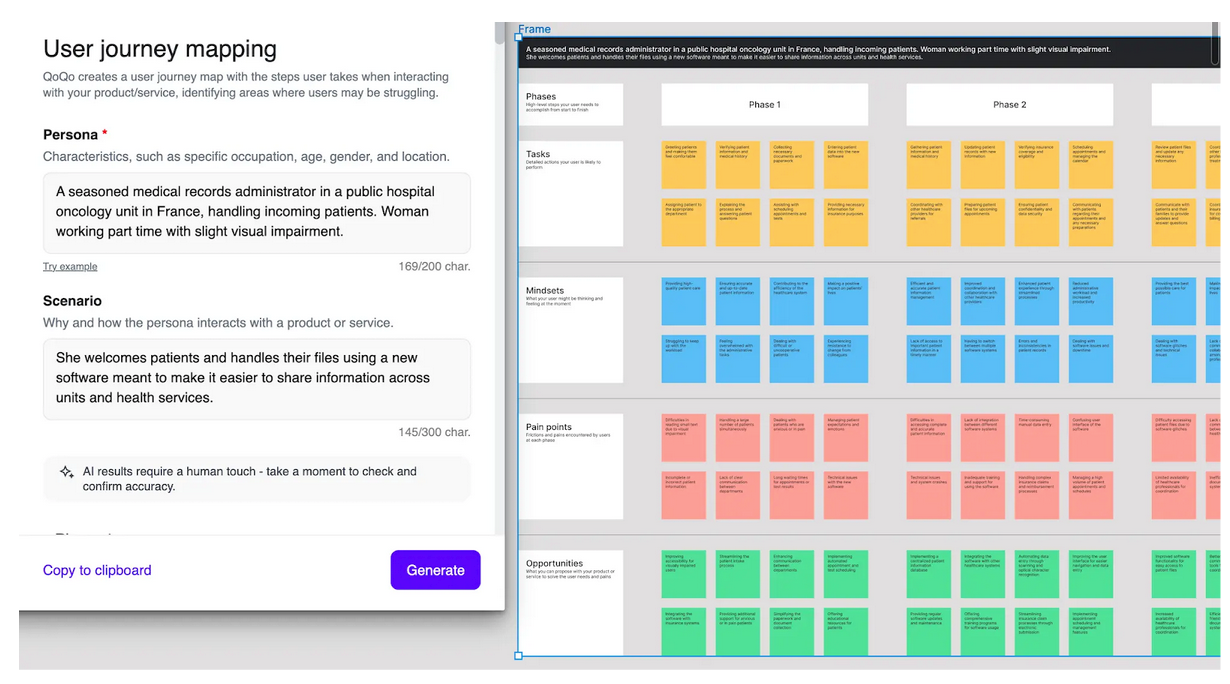}
  \caption{\label{fig:img5}Screenshot of the UI for the generation  of the User Journey Mapping in the service QoQo}
\end{figure}

Moreover, in these services it is possible to automate the definition of journey mapping, a model  that can be used to visualise the process that a person goes through in order to achieve a specific goal. QoQo (see Figure \ref{fig:img5}), for example, takes as input a persona and a scenario describing the context in which the interaction takes place; as a result the system fills in a table, articulated in different phases, with corresponding tasks, mindsets, pain points, and opportunities.

Although these tools are not developed specifically for cultural heritage, their validity in this domain can be proved by comparing these results with CH-specific design toolkits. As an example, the aforementioned \textit{VisitorBox} facilitates the collaboration and the co-creation between designers and practitioners, supporting them throughout the entire design workflow, from the definition of requirements to the ideation of the final experience. Therefore, it also accompanies them in the analysis of the target audience, which is structured in a similar way to the AI-powered services cited in this section: designers are indeed asked to indicate possible motivations, barriers, capabilities and devices which may be used during the interaction, which reflects the pieces of information included in user personas and journey maps.

\subsection{Phase 3: Design}
\label{a:phase3}

As illustrated in the description of the workflow, the third stage constitutes the core of the design process: here, the requirements derived from the analysis of the state of the art and from the PACT Framework are organised in the design brief, which is the technical document upon which all the subsequent stages of this workflow are based (see \textit{supra}). In addition, the brief may be accompanied by mock-ups and wireframes. Moving forward, the following section presents a small selection of AI-powered tools which exemplifies how Artificial Intelligence techniques may be implemented with this objective.

\subsubsection{Task 3.1: Definition of the design brief}
\label{a:phase3-1}

Natural Language Generation techniques provide useful solutions to accomplish this task, and platforms like QoQo or GoodBrief (\url{https://goodbrief.io/}) are effectively the perfect means to solve this task. Nevertheless, among the available solutions, the Design Brief Generator by AI Planet (\url{https://aiplanet.com/apps/design-brief-genrator}) is one of the most interesting as it can produce convincing results with few input parameters and can be accessed for free. Here, the user is asked to provide a synthetic description of the target audience and of the most relevant features of the final application. As a response, an AI system has been trained to return a full design brief articulated in seven different sections: overview, target audience, objective, scope of work, deliverable, timeline, and budgets.

This platform has been tested on one of the experiments currently being developed at CNR ISPC: \textit{BrancacciPOV}. It consists of a virtual tour, which is designed to support a guided exploration of the Brancacci Chapel in Florence remotely and with multiple users, accessible through different devices (see \cite{brancaccipov}). With an extremely synthetic input (“general public in a cultural site” as target audience, while its features have been summarised as follows: “the interactive application allows visitors to enjoy a virtual collaborative tour of the cultural site”), a promising outcome is returned.
The provided inputs are too general and the resulting brief sometimes lacks specificity. However, the output of AI Planet manages to provide interesting suggestions: according to the AI-generated document, the workflow should include at the very beginning a comprehensive analysis of the cultural site. In addition, the UX should enable visitors to choose between different languages and ensure compatibility with different devices: these elements play indeed a crucial role in the current version of the application, which is available both in English and Italian and has been developed with the ATON, a web-based XR framework suitable for mobile devices, laptop and VR headsets \cite{fanini2021aton}.

Furthermore, the retrieved brief suggests the implementation of “social features that allow visitors to interact and collaborate during the virtual tour, such as chat functionality and sharing of photos and videos”. This feature is currently not available in the current version of \textit{BrancacciPOV}, but it may be included in the redesign of the UX in the next releases. In this context,  the role of AI as a “co-pilot", which was introduced in the definition of the research question, becomes clear, as it can support the designers by providing additional suggestions and inspirations for the improvement of the final application.

\subsubsection{Task 3.2: Definition of mockups and wireframes} \label{a:phase3-2}

\begin{figure}[htb]
  \centering
  \includegraphics[width=.8\linewidth]{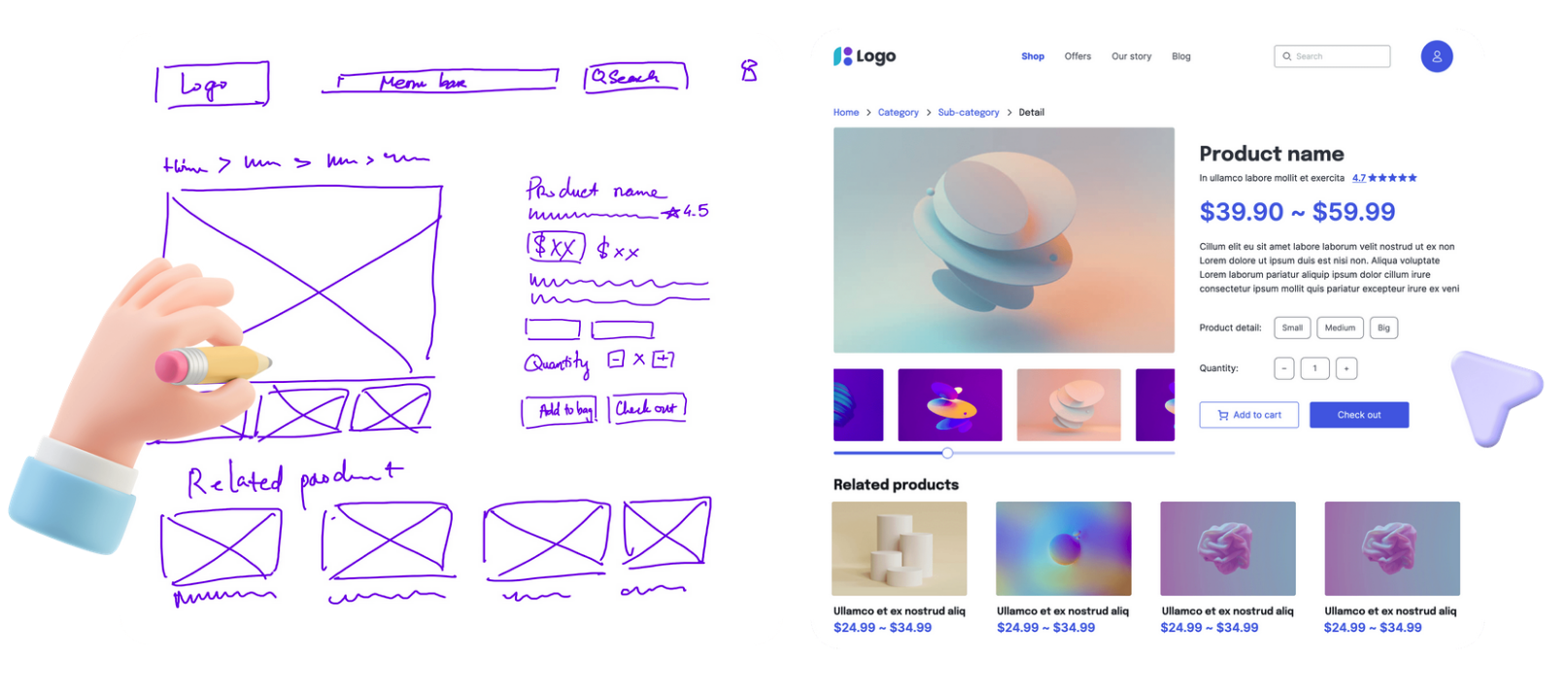}
  \caption{\label{fig:img6}Screenshot of the input and output of the service of AI Design (wireframe conversion) in  Visily.}
\end{figure}

Ad hoc services are trained to streamline the process of wireframe and prototype creation. The platform Visily (\url{https://www.visily.ai/}) reuses the metaphor of the “AI codesigner” since it is able to perform grammar check and tone detection, as well as to look for similar images. In addition, it provides a service to convert wireframes from lo-fi into hi-fi version. A beta version is currently under development which should be able to create a first prototype starting from a textual input (as shown in Figure \ref{fig:img6}). Similar functionalities can be performed by other platforms as well, such as Uizard (\url{https://uizard.io/}) and UI-AI (\url{https://ui-ai.com/}).

In the last few years, AI tools have also been introduced in Figma (\url{https://www.figma.com/}), one of the best known prototyping environments: here, traditional prototyping activities such as image and icon generation can be delegated to AI tools, as with the plug-in Magician (\url{https://magician.design/}). Additionally, the plug-in “html to design” (\url{https://html.to.design/home}) enables the direct conversion of a website into an editable Figma prototype.

\subsection{Phase 4: From the Design of the Solutions to the Creation of Prototypes} \label{a:genai}

As anticipated in Section \ref{a:phase3-2}, traditional prototyping can be tackled by GenAI systems. This class of algorithms is a powerful support for the development of the elements of minimum viable products, which are the core of the fourth stage of the current workflow. As a result, this section will analyse these two stages together, providing a holistic overview of possible GenAI tools, distinguishing them  according to the format of their outcome. More specifically, major attention will be paid to the generation of natural and artificial language (i.e. code development), of audio (both music and voice with synthesisers), of images and other visual content (in particular, video and 3D assets). The generation of data (i.e. synthetic data) will instead be presented during the analysis of AI-based testing solutions (see Section \ref{a:phase5-test}).

\subsubsection{Natural Language Generation, Translation and Storytelling}
\label{a:genai-nlp}

The impact of ChatGPT and other prompt-based chatbot is undeniable and more and more designers are implementing these tools in their everyday practice. However, other platforms can offer more specific solutions for the creation and management of textual contents which are necessary for the preparation of an interactive experience for CH.

Sudowrite (\url{https://www.sudowrite.com/}) is thought as a co-brainstorming buddy: it is able to suggest a possible continuation for a given paragraph, to generate description of an element mentioned in the narration, as well as other functions such as tone shifting. Conversely, Jasper (\url{https://www.jasper.ai/}) extends the capabilities of ChatGPT by implementing the possibility to add company-specific documents to the knowledge base. From this data the AI service is able to create text based on pre-trained templates, specific for marketing, social media management and content strategy.

PlotFactory (\url{https://plotfactory.com/}), on the other hand, does not exclusively support text generation, which can be defined as an AI-assisted story editor. Users can rely on ready-made templates and text-to-speech services; in addition, it is designed to help users organise chapters and manage different elements of the narration \cite{palombiniStorytellingTellingHistory2017}. In particular, it is able to keep track of places, characters, and relevant objects. This tool, mainly designed for creative writing tasks, can be implemented in the preparation of specific products in the domain of CH: for example, serious games require a complex storytelling structure, composed of different elements (e.g. non-playable characters, scenarios, items), which can benefit from the use of a similar platform.

Another area of application is automatic translation systems, a traditional domain of NLP, which can assist support designers and practitioners in managing multilingual workflows. Wang and colleagues already provided a systematic overview of this topic \cite{WANG2022143}: in their study, they trace the history of this field of research, from its first proposal by Warren Weaver in 1947 to the most recent flagship technologies such as Recurrent Neural Networks and Transformer architectures. While acknowledging that there is “room for improvement", the authors highlight the remarkable progress made in machine translation over the past decades in several domains, including CH. Currently, online services can offer fast and good quality translations, such as ChatGPT itself and DeepL (\url{https://www.deepl.com/it/translator}).

\subsubsection{Image Generation} \label{a:genai-img}
Another domain in which the implementation of GenAI is increasingly gaining popularity is image generation. In particular, in the last decade, AI performances in Computer Vision (CV) has undeniably improved and the advent of text-to-image algorithms has also enabled the general public to create visual content without specific artistic or design skills. Cao and colleagues \cite{caoComprehensiveSurveyAIGenerated2023} identified a pivotal moment in the introduction of transformer-based architectures in 2017, as shown in the timeline of Figure \ref{fig:img7}:

\begin{figure}[htb]
  \centering
  \includegraphics[width=.8\linewidth]{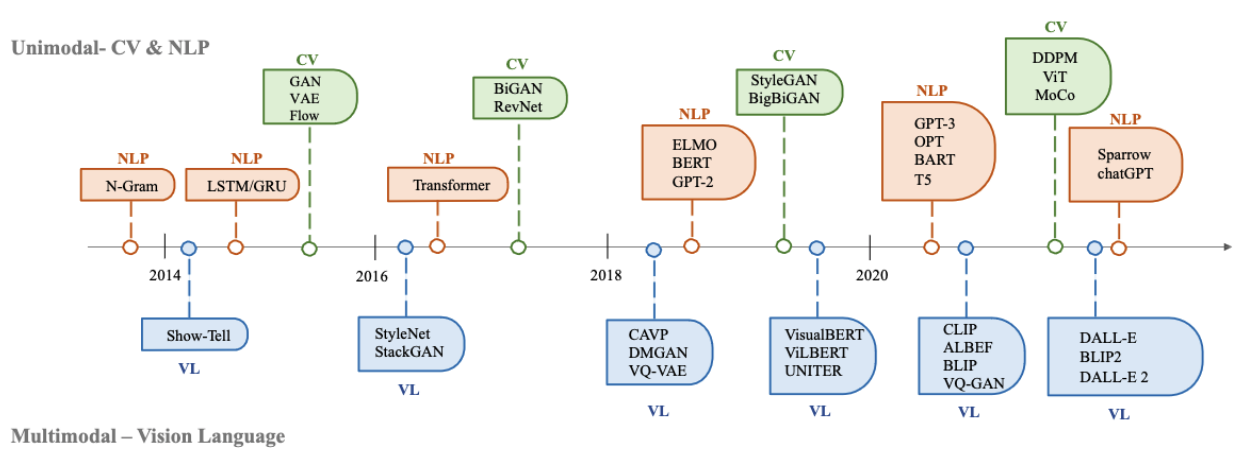}
  \caption{\label{fig:img7}The history of Generative AI in CV, NLP and Vision Language (VL) \cite{caoComprehensiveSurveyAIGenerated2023}.}
\end{figure}

Traditional text-to-image genAI relies on two different architectures, which are well exemplified by two of the most controversial images in the history of digital arts (for a more detailed analysis see \cite{pedrazziFuturiPossibiliScenari2021}): \textit{Portrait of Edmond de Bélamy} (2018) and \textit{Théatre d’Opéra Spatial} (2022,\cite{book}).

The first canvas, created by the collective Obvious, became famous when it was sold at a Christie’s auction for 432.5 thousands euros. It was generated through GANs, an architecture firstly introduced in 2014, which is articulated in two core structures: a discriminator and a generator. The former is trained to distinguish elements from a dataset, whereas the latter is trained to produce images that the discriminator cannot differentiate from the original data. The aim is for the generator to create realistic images that are mistaken for genuine portraits.

Conversely, the second canvas was presented by the artist Jason Allen and won the Colorado State Fair Fine Arts Competition. The used GenAI, Midjourney, relied on diffusion models: here, some Gaussian noise is added to a picture. Subsequently, the model denoises the resulting blurry picture step by step in order to finally reconstruct a picture  that resembles the original one as much as possible. At the same time, a textual input (a description) guides the process of denoising (for a “gentle” introduction to GAN and diffusion models, see \cite{book}).

Different platforms now offer similar services, such as OpenAI’s  DALL•E (\url{https://openai.com/dall-e-3}), together with Midjourney (\url{https://www.midjourney.com/home)}, Dreamstudio (\url{https://dreamstudio.ai/}) and Getty Images (\url{https://www.gettyimages.co.uk/ai/generation/about}). In addition, Adobe, one of the most commonly used suites for designers, recently developed an AI-powered image generation functionality, Adobe Firefly (\url{https://www.adobe.com/it/products/firefly.html}), already embedded in some of its most popular applications, such as Adobe Photoshop. 

\subsubsection{3D Assets Generation} \label{a:genai-3d} 

Since 2020, Computer Vision research has been witnessing better and better performances in the generations of the 3D objects through AI models \cite{gozalo-brizuelaChatGPTNotAll2023}. These methods are often subdivided according to the accepted input:

\begin{itemize}
    \item Textual input, such as Google’s Dreamfusion \cite{pooleDreamFusionTextto3DUsing2022}, OpenAI’s Point•E \cite{nicholPointESystemGenerating2022}, Apple’s GAUDI \cite{bautistaGAUDINeuralArchitect2022}, NVIDIA’s Magic3D and MVDream by ByteDance and University of California, St. Diego \cite{shiMVDreamMultiviewDiffusion2023}; 
    \item Visual input: Although different experiments have been already carried out in the domain of 3D modelling from 2D images, a key study in 2020 marked a new state-of-the-art technique. Using a series of images (or a video) as input, Neural Radiance Fields (NeRF) are actually able to build a static 3D scene \cite{mildenhallNeRFRepresentingScenes2020}. Starting from this solution, other models have been developed, such as NeROIC \cite{kuangNeROICNeuralRendering2022} and SMERF \cite{duckworthSMERFStreamableMemory2024}, yet the debate on this class of genAI is still open \cite{remondinoCriticalAnalysisNeRFBased2023}.
\end{itemize}

The rise of research products in this domain resulted in an increasing availability of web-services enabling non-specialist users to instantly create 3D objects relying on genAI models. Some possible examples of these platforms are listed in Table \ref{tab:3D} (integration to \cite{wodeckiAI3DGeneration2023}, please note that Meshcapade also accepts other data formats as input, such as measurements and scans, and can transform videos into full animations).

\begin{table}[h]
 \caption{A first overview of genAI-powered web services for the creation of 3D Assets}
  \label{tab:3D}
\centering
\begin{tabular}{>{\hspace{0pt}}m{0.747\linewidth}>{\centering\hspace{0pt}}m{0.061\linewidth}>{\centering\arraybackslash\hspace{0pt}}m{0.129\linewidth}}
                                                                                                                                                                                                                                   &               &                                                    \\
\multirow{2}{0.747\linewidth}{\hspace{0pt}Service}                                                                                                                                                                        & \multicolumn{2}{>{\hspace{0pt}}m{0.19\linewidth}}{Input}  \\
                                                                                                                                                                                                                                   & \textit{Text} & \textit{2D Image}                                  \\ 
\hline
Meschapade ({\url{https://meshcapade.com/}}) &               & $\checkmark$                                                  \\
Masterpiece ({\url{https://www.masterpiecex.com/}})                                                                                                                                              & $\checkmark$             &                                                    \\
Ponzu ({\url{https://www.ponzu.gg/}})                                                                                                                                                            & $\checkmark$             &                                                    \\
Spline ({\url{https://spline.design/ai}})                                                                                                                                                        & $\checkmark$             &                                                    \\
3DFY.AI ({\url{https://app.3dfy.ai/}})                                                                                                                                                           & $\checkmark$             &                                                    \\
Sloyd ({\url{https://www.sloyd.ai/}})                                                                                                                                                            & $\checkmark$             &                                                    \\
Gliastar ({\url{https://lumalabs.ai/}})                                                                                                                                                          &               & $\checkmark$                                                  \\
Fotor ({\url{https://www.fotor.com/features/ai-3d-model-generator/}})                                                                                                                            & $\checkmark$             &                                                    \\
Alpha3D ({\url{https://www.alpha3d.io/}})                                                                                                                                                        &               & $\checkmark$                                                  \\
Luma AI ({\url{https://lumalabs.ai/}})                                                                                                                                                           & $\checkmark$             &                                                   
\end{tabular}
\end{table}

These services, however, are not specifically developed for the CH sector. As a matter of fact, they often deal with simple objects, such as pieces of furniture or props (for instance, to be used in video games), or are specifically thought for the modelling of 3D humanoid characters (as in Meshcapade, Masterpiece, Gliastar, and Fotor) and their animations from video input (besides Meshcapade, see also DeepMotion, \url{https://www.deepmotion.com/animate-3d}, and RokokoVision, \url{https://www.rokoko.com/products/vision}). On the contrary, their performance with tridimensional artworks (such as sculptures) is unsatisfactory. As a result, given the current state of the art in this domain, a genAI technique that can streamline reality-based acquisition techniques (e.g. laser scanner and photogrammetry) is still missing. 

\subsubsection{Audio and Audiovisual Content Generation} \label{a:genai-mp3-4}
Various studies show an improvement in the final UX thanks to the implementation of audio elements \cite{kellingImplicationsAudioNarration2018}, however the creation of this kind of content may be particularly time-consuming. AI systems, on the other hand, can provide satisfactory solutions for speech synthesis, and music and video generation.

As a matter of fact, several text-to-speech services are available online. Based on different architectures, such as the encoder-decoder structure or the latest transformer-based solutions (see a more precise description in \cite{tamburini2022neural}), they are able to generate an audio starting from a textual input, which can be rendered in a variety of voices. Some examples are IBM (\url{https://www.ibm.com/products/text-to-speech}), MurfAI (\url{https://murf.ai/)}, Lovo (\url{https://lovo.ai/}), Synthesis (\url{https://synthesys.io/}), Speechify (\url{https://speechify.com/}), Verbatik (\url{https://verbatik.com/}).

In recent years, DL algorithms have enabled the rapid creation of music production also for users with no prior knowledge in composition. These models are trained on large datasets to recognise patterns within pieces and to generate compositions around these patterns, starting from a huge variety of inputs (tempo, key, genre, melodies, harmonies, rhythms). One of the best known services is AIVA (\url{https://www.aiva.ai/}), which can take as input a style preset or a chord progression and allows users to further edit to the final result. Direct editing on the online platform is also implemented in Soundful (\url{https://soundful.com/}), while Ecrett Music (\url{https://ecrettmusic.com/}) accepts a variety of inputs from a selection of scenes, moods, and genres. In addition, Suno AI (\url{https://suno.com/}) enables users to create full song starting from a textual input. 

Also the creation of video can be delegated to different platforms. For instance, FlexClip (\url{https://www.flexclip.com/}), DeepBrainAI (\url{https://www.deepbrain.io/}), and InvideoAI (\url{https://invideo.io/}) are able to generate videos from a textual input. Other platforms, such as synesthesia (\url{https://www.synthesia.io/}) and Pictory (\url{https://pictory.ai/}), integrate additional post-production effects, through which users may add subtitles and select within a significant range of video editing tools.

One of the latest innovations in this field is produced by a flagship company in this sector, OpenAI, which released Sora (\url{https://openai.com/sora}) in mid February 2024. It has been presented as a “generalist model of visual data”, capable of generating “videos and images spanning diverse durations, aspect ratios and resolutions, up to a full minute of high definition video” \cite{videoworldsimulators2024}. It consists of a diffusion model, based on a transformer architecture and trained on patches of visual data, working in synergy with DALL•E 3 and GPT.

Consequently, Sora is capable of generating video starting from a variety of inputs: although it is primarily conceived as a text-to-video GenAI, “Sora can also be prompted with other inputs, such as pre-existing images or video”, which “enables [it] to perform a wide range of image and video editing tasks—creating perfectly looping video, animating static images, extending videos forwards or backwards in time”, and editing video (e.g. change the setting). Because of its recent release, no examples of implementation are available at the time of writing - especially in the CH sector -, yet the performances described in Open AI technical report \cite{videoworldsimulators2024} promise to greatly innovate in this area.

\subsubsection{AI Support in Code Development} \label{a:genai-code}
Transformer-based chatbots, such as ChatGPT, already performed well in the domain of code generation: a recent study observes how ChatGPT “is used in a variety of code generation domains, e.g., debugging codes, preparing programming interviews, and solving academic assignment” \cite{feng2023investigating}. In these systems, code snippets are generated by prompt-based neural architectures, which have been trained on vast datasets of different source codes from open-source projects. Programmers enter plain text prompts describing what they want the code to do and can also have the code “translated” from one language to another.

Beside ChatGPT, other services can perform this task, such as Tabnine (\url{https://www.tabnine.com/}), IBM Watson Code Assistant (\url{https://www.ibm.com/products/watsonx-code-assistant}) and GitHub Copilot (\url{https://github.com/features/copilot}): the latter is the fruit of the collaboration between GitHub and OpenAI, both owned - at least in part - by Microsoft. It is a ChatGPT-like chatbot, trained on a GPT-3 architecture and directly embedded in an integrated development environment (IDE, e.g. Visual Studio Code). Users can interrogate this system to create new snippets of code or better understand already existing algorithms, which may be written in a variety of languages such as Python, JavaScript, TypeScript, or Ruby.

In addition, more specific solutions for web development have been put forward. Similarly to the platforms defined in Section \ref{a:phase3-2}, certain systems are able to convert prototypes into the source HTML structure and stylesheets: this task is partially carried out by the Figma Dev Mode (\url{https://www.figma.com/dev-mode/}), thanks to which can generate production-ready CSS, iOS, or Android code snippets from a prototype. More complete results were instead reported in a study by Asirolgu and colleagues, who compared different algorithms for converting mockups and wireframes into an HTML document using Deep Learning: this analysis showed that a 2019 project \cite{asirogluAutomaticHTMLCode2019} achieved the benchmark of 96\% method accuracy (see a more detailed analysis in the reference study \cite{daveSurveyArtificialIntelligence2021}).

\subsection{Phase 5: Testing}
\label{a:phase5-test}
\begin{figure}[htb]
  \centering
  \includegraphics[width=.8\linewidth]{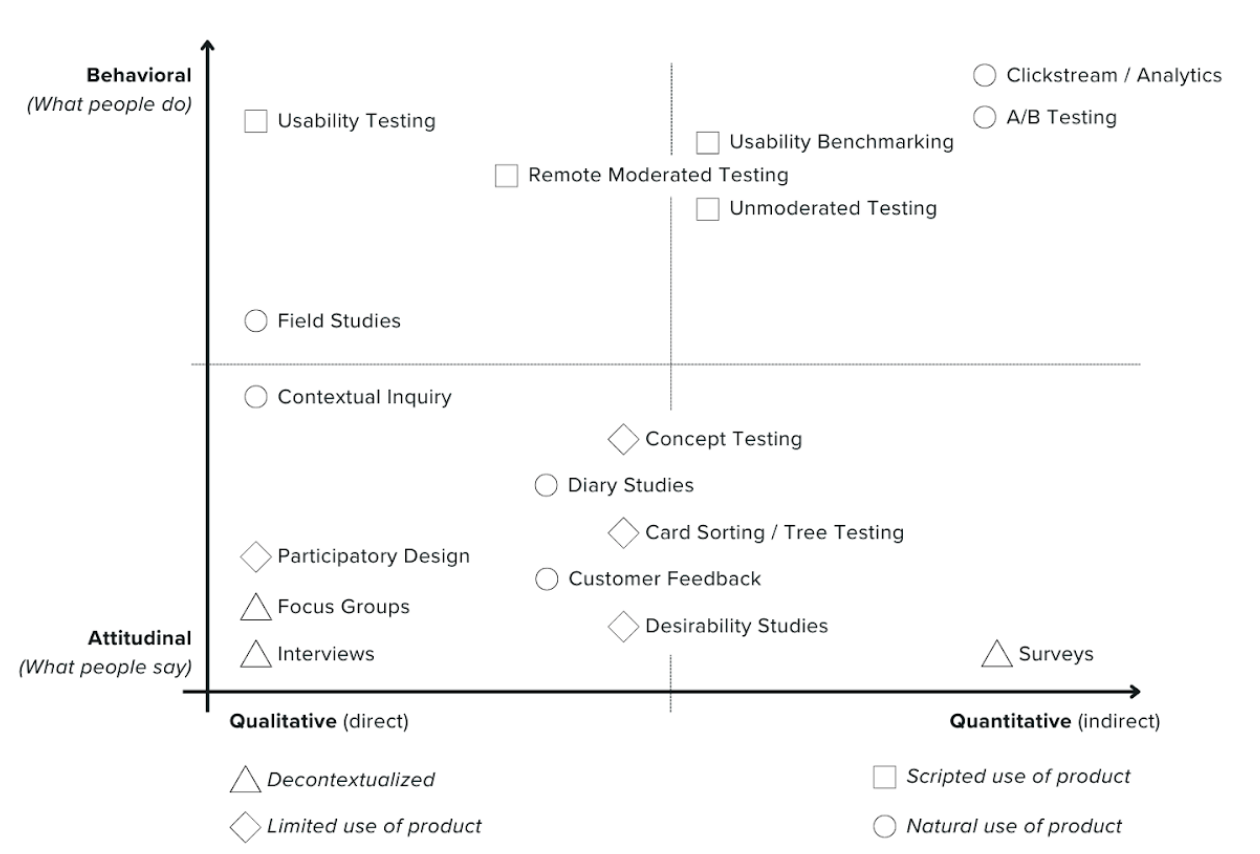}
  \caption{\label{fig:img11}Landscape of user research methods (after  \cite{rohrerWhenUseWhich2022}}
\end{figure}

The evaluation of the visitor experience is a crucial step in the IxD workflow: at this stage, designers pay attention to the achievement of the cognitive goal, together with the traditional usability assessment. This latter category of tests entails a wide variety of evaluation methods, which have been systematised in a convincing chart by Christian Rohrer, as shown in Figure \ref{fig:img11} \cite{rohrerWhenUseWhich2022}. The researcher subdivides these methods according to the focus and the modality of the assessments: the first parameter discriminates between what people think (\textit{attitudinal} tests) and what they do (\textit{behavioural} tests). Conversely, the latter axis differentiates between qualitative and quantitative methods.

Focussing on this latter subdivision is useful also when trying to understand the multifaceted landscape of AI-powered UX assessment tools. Qualitative methods, such as interviews, often rely on free-text answers: consequently, these services exploit NLP techniques to retrieve the most meaningful pieces of information from the interviewed audience, for example through summarization, topic modelling, polarity detection, sentiment analysis. On the other hand, quantitative methods require a different analytical approach, which are mainly based on the creation of frequency matrices and systems of data visualisation.

Among all the available platforms, UserTesting (\url{https://www.usertesting.com/}), which now includes also UserZoom, is conceived as a service supporting designers in all the different steps of the UX evaluation, from audience recruitment to preparation of different typologies of tests. It also implements AI to retrieve user analytics, through which friction points or interactive path flow may be individuated. AppliTools (\url{https://applitools.com/}), instead, offers tailored solutions for the evaluation of user interfaces and provides automated tests of the functional requirements and responsiveness check. Lastly, Maze (\url{https://maze.co/}) features additional services for UX/UI, such as prototype testing with real users and interviews. It also uses AI powered tools through which it is possible to create a personal survey and let AI simulate answers and perform sentiment analysis and interviews summaries.

In this landscape, one of the most original solutions is the one proposed by muse (\url{https://muse.stream/en/}) developed by Sarah Kenderdine, director of the Laboratory of Experimental Museology (eM+) at EPFL Lausanne \cite{kocsiskenderdine}. Its peculiarity lies in the fact that it is designed explicitly for the assessment of visitors’ experience in cultural institutions. It may resemble other survey platforms, such as Google Form, yet the variety of the available question formats is far richer and specific ML algorithms are able to provide more precise visualisations of the collected data.

A final promising intersection between AI services and UX testing methods is the implementation of synthetic data. They indeed consist of computer-generated data which accurately represent and augment a real-world dataset: they are hence a powerful tool which could potentially streamline the process of data collection. Although the existing literature and platforms (see Synthesized, \url{https://www.synthesized.io/}) mainly mentions their implementation in software and AI engineering (e.g. training of ML models), an innovative solution for UX design is being proposed by Synthetic Users (\url{https://www.syntheticusers.com/}).

This platform enables designers to perform UX research without a test with real users. The AI architecture, based on OpenAI GPT 3.5 and 4 models, takes as inputs a persona’s characteristics and its possible problems with a proposed solution: as a result, it will be able to run virtual user interviews, gathering feedback from the synthetic personas and summarising it in a complete report. Again, in the FAQ section of their website, the metaphor of the co-pilot is reused, since this service is claimed to have ”been designed to act as a discovery co-pilot”, accelerating “an otherwise expensive and operationally taxing process”. These synthetic users have been tested via a comparison with organic interviews and revealed the reliability of this platform \cite{syntheticusersHowWeMeasure2023}. Lastly, the promising results of GAN-generated synthetic data in UX research methods have been proven also in the academic fields, in particular with regards to UI assessment \cite{Jalal1772838}.

\section{Discussion}
\label{a:discussion}
\begin{figure}[htb]
  \centering
  \includegraphics[width=.8\linewidth]{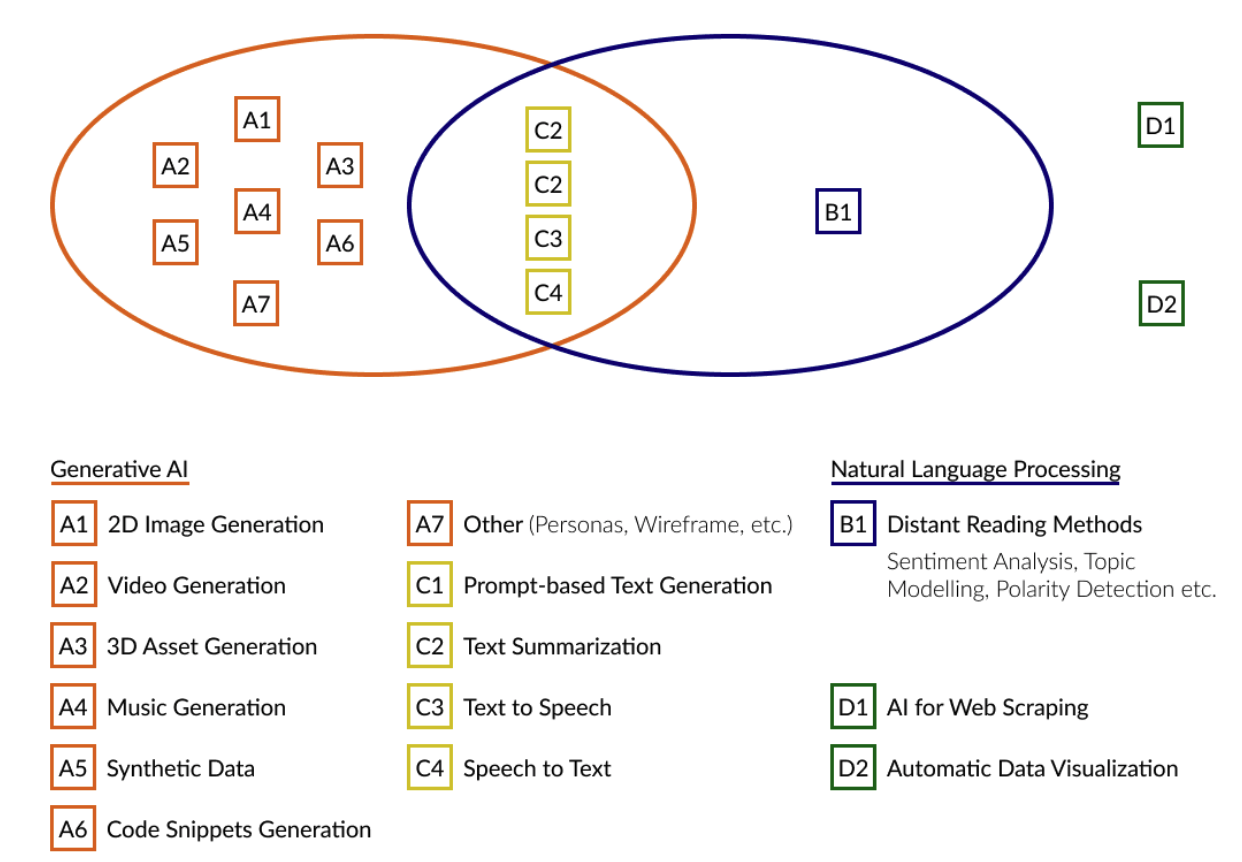}
  \caption{\label{fig:img12}Synoptic view of the classes of tools and services presented in this study}
\end{figure}

The technologies presented in this paper often fall into two major categories, i.e Natural Language Processing and Generative AI, in particular for textual and image content. They indeed coincide with two flagship fields of application of Machine and Deep Learning algorithms, which have witnessed an unprecedented attention also outside academia (as in the case of OpenAI service ChatGPT). As highlighted throughout this research, these domains are not completely disjoint, rather their intersection entails services such as chatbots or text summarizations tools. To this regard, Figure \ref{fig:img12} offers a synoptic overview of the different tools and services presented in this article, proposing a possible taxonomy. 

At the light of this consideration, the current possibilities for the implementation of AI techniques are not homogeneous throughout the different stages of the workflow described in Section \ref{a:workflow}. As stated in the methodological note of this study, this state of the art overview did not perform a quantitative assessment of the available tools. As a result, Table \ref{tab:discussion} proposes a first qualitative overview of the different performance of AI systems with reference to the tasks mentioned in Section \ref{a:list-resources} and with the categories listed in Figure \ref{fig:img12} .

\begin{table}
 \caption{Synoptic overview of the classes of tools and services presented in this study with the corresponding task}
  \label{tab:discussion}
\centering
\begin{tblr}{
  width = \linewidth,
  colspec = {Q[184]Q[313]Q[100]Q[100]Q[109]Q[100]Q[25]},
  cell{1}{1} = {r=2}{},
  cell{1}{2} = {r=2}{},
  cell{1}{3} = {c=4}{0.409\linewidth},
  cell{3}{1} = {r=3}{t},
  cell{6}{1} = {r=2}{t},
  cell{8}{1} = {r=3}{t},
  cell{11}{1} = {r=3}{t},
  cell{14}{1} = {r=2}{t},
  hline{3} = {1-6}{},
}
Design Phase          & Task                        & Technological support &                           &                           &                           &  \\
                                &                                      & \textit{Dom. A}      & \textit{Dom. B} & \textit{Dom. C} & \textit{Dom. D} &  \\
1. Inspirations                 & 1.1 Analysis of the CH object / Site     &                                &                           & C1, C2                    & D1                        &  \\
                                & 1.1 State of the art - Literature review &                                &                           & C1, C2                    & D1                        &  \\
                                & 1.2 State of the art - Document analysis &                                &                           & C1, C2                    &                           &  \\
2. PACT Framework               & 2.1 People - User Understanding~         &                                & B1                        & C2, C4                    &                           &  \\
                                & 2.2 People - User Modelling              & A7                             &                           & C1                        &                           &  \\
3. Design                       & 3.1 Design Brief Preparation             &                                &                           & C1                        &                           &  \\
                                & 3.2 Wireframing                          & A7                             &                           &                           &                           &  \\
                                & 3.3 Design of the solution               & {A1, A2,\\A3, A4  }            &                           & C1, C2, C3                &                           &  \\
4. Prototyping                  & 4.1 Asset Creation                       & {A1, A2,\\A3, A4  }            &                           & C1, C2, C3                &                           &  \\
                                & 4.2 Content Creation                     & {A1, A2,\\A3, A4  }            &                           & C1, C2, C3                &                           &  \\
                                & 4.3 Solution implementation~             & A6                             &                           &                           &                           &  \\
5. Testing (\textit{iterative}) & 5.1 Data Retrieval                       & A5                             &                           & C1                        &                           &  \\
                                & 5.2 Data Analysis                        &                                & B1                        &                           & D2                        &  
\end{tblr}
\end{table}

The table shows the heterogeneity of the implementation of AI systems classes in IxD workflows and highlights the relevance of GenAI and NLP. Only other two types of Artificial Intelligence are indeed excluded from these two sets and have a marginal field of application: web scraping tools and automatic data visualisation (labelled with the letter “D” in Figure \ref{fig:img12}  and Table \ref{tab:discussion}). On the contrary, the use of generative algorithms for the creation of natural language is by far the most common and is implemented transversally throughout the entire workflow.

The rise and the widespread use of generative systems also by the general public had a major impact also on the potential application in creating cultural experiences. Their performance, however, is considerably different: whereas text-to-image algorithms are now able to reach satisfying results and have been already incorporated in traditional design software suites (e.g. Adobe), other models do not provide a significant improvement compared to traditional methods. This is particularly evident in the creation of 3D models, whose quality still needs to be improved, given that AI algorithms still fail in providing reliable reproductions of cultural heritage objects.

To what pertains to the domain of UX evaluation, the scope of use of synthetic data is still limited and needs to be better explored in the future. On the contrary, the majority of tools in the last phase of the workflow are used for the automatic analysis of users’ responses and analytics, either for quantitative responses (category D2) or for traditional “distant reading” methods \cite{moretti2013distant}, such as sentiment analysis, topic modelling and polarity detection , which have been proven effective in the analysis of corpora of textual data. 

\section{Conclusion}
\label{a:conclusion}
This paper presents a first overview on the fields of implementations of Artificial Intelligence in the development of interactive visitor experience for Digital Cultural Heritage. This research aims hence at addressing an existing gap in the literature, since - to the best of my knowledge - no study has been conducted maintaining the specificities of both UX and CH domain. As demonstrated in Section \ref{a:sota}, previous publications either considered the single stages of the UX workflow without explicitly referring to the peculiarities of CH sector \cite{stigeArtificialIntelligenceAI2023}; on the other hand, analyses on the possible applications of AI for the enhancement of visitor experience in cultural institutions does not refer to the best practices and methodologies of UX and IxD design \cite{pisoniHumanCenteredArtificialIntelligence2021}.

In addition, existing overviews of the state of the art are based on literature reviews of scholarly publications, which may ignore proprietary services, whose release may not be accompanied by articles or technical papers. Section \ref{a:methodology} hence describes the methodological basis of this study, which has been carried out following deductive approach. At first a workflow for IxD in CH is proposed, which is articulated in phases and tasks: for each of them, a possible AI-based technique is proposed. The services mentioned in this paper do not make up an extensive catalogue and the inclusion or exclusion of a specific product should not be interpreted as an implicit judgement of its performance. On the contrary, they are only an exemplification of a possible use case of an AI technique for a specific task in the IxD workflow.

After a rapid overview on current reuse of AI tools by UX designer, the core section of this paper analyses different AI-services to automatise the tasks mentioned in the workflow described in Section \ref{a:workflow}. This review mainly included solutions which may be used by users without coding capabilities and showed that almost the totality of the tasks may benefit from the implementation of at least one AI-based approach. The services and algorithms presented in Section \ref{a:list-resources} greatly differ in terms of purposes and architectures, yet two major classes of techniques may be identified. Since several UX research methods are based on the gathering of qualitative data, in particular in textual format, NLP techniques are frequently used in this domain. In addition, GenAI too emerges as a predominant class of techniques which may enhance IxD workflow in this sector.

In general, these solutions are often presented as a support tool for designers and not as a definitive replacement. This is evident in the recurrent use of metaphors such as “copilot” or “co-brainstorming buddy”, which underlines the potentialities of these methods: indeed, in design AI manages to both automatise and streamline mechanical and time-consuming task and to propose novel ideas and suggestion to enhance the final product. In this scenario, however, it is possible to notice that some specific tasks are being paid more attention than others. Whereas NLP-based platforms (e.g chatbots, summarization, translation and text analysis tools) are now common, more recent techniques, such as AI-driven 3D modelling and synthetic data for UX/UI evaluation, are either more rare or their performance needs to be improved.

In lieu of conclusion, this research tried to inquire about the possibilities offered by the multifaceted landscape of AI-applications, in particular to designers, curators and cultural heritage professionals, who may not have solid programming skills. This study, however, briefly captures the current state of the art, which however is expected to greatly change in the close future, as in the case of OpenAI’s Sora and text-to-video genAI. In addition, no proper distinction has been made in this research between proprietary and open-source platforms: future research in this domain might thus concentrate on which are the possibilities offered by the latter class of solutions. 

Finally, this study limits its focus to the most relevant tasks of a currently established IxD workflow and does not consider the potential AI-driven transformations of the design methodology as a whole in the near future. Studies and evaluations are therefore needed to investigate the impact of artificial intelligence systems in visualising and ideating entire visitor experiences from a set of requirements, rather than just performing individual tasks. The results of this research would shed new light on the extent to which these tools can be considered a reliable co-creative agent. In this perspective, further research should also be devoted to analysing the ethical implications of integrating AI in this domain, taking into account the complexity and open questions related to Digital Cultural Heritage.

\bibliographystyle{unsrt}  
\bibliography{references}  


\end{document}